\documentclass[twocolumn,english]{IEEEtran}
\usepackage[T1]{fontenc}
\usepackage[utf8]{luainputenc}
\usepackage{amsmath}
\usepackage{amssymb}
\usepackage{graphicx}

\makeatletter

\ifCLASSOPTIONcompsoc
\else
\usepackage{multicol}\fi

\usepackage{babel}

\makeatother

\usepackage{babel}
\begin{document}

\title{On the Performance of Multiple Antenna Cooperative Spectrum Sharing
Protocol under Nakagami-$m$ Fading%
\thanks{Part of this work was supported by Deity (Department of Electronics
and Information Technology), Government of India, under SRP-43 project
grant. %
}}

\author{\begin{multicols}{3} Shubha Sharma\\
Wirocomm Research Group\\
 Dept. of Electronics and\\
 Communication Engg.\\
Indraprastha Institute of \\
Information \& Technology\\
 (IIIT-Delhi), New Delhi, India \\
shubha12104@iiitd.ac.in\\
 Ankush Vashistha\\
 Wirocomm Research Group\\
Dept. of Electronics and\\
 Communication Engg.\\
 Indraprastha Institute of \\
Information \& Technology\\
 (IIIT-Delhi), New Delhi, India \\
ankushv@iiitd.ac.in\\
 Vivek Ashok Bohara\\
Wirocomm Research Group\\
 Dept. of Electronics and\\
 Communication Engg.\\
 Indraprastha Institute of \\
Information \& Technology\\
 (IIIT-Delhi), New Delhi, India \\
vivek.b@iiitd.ac.in\\
 \end{multicols}}
\maketitle
\begin{abstract}
In a cooperative spectrum sharing (CSS) protocol, two wireless systems
operate over the same frequency band albeit with different priorities.
The secondary (or cognitive) system which has a lower priority, helps
the higher priority primary system to achieve its target rate by acting
as a relay and allocating a fraction of its power to forward the primary
signal. The secondary system in return is benefited by transmitting
its own data on primary system's spectrum. In this paper, we have
analyzed the performance of multiple antenna cooperative spectrum
sharing protocol under Nakagami-m Fading. Closed form expressions
for outage probability have been obtained by varying the parameters
$m$ and $\Omega$ of the Nakagami-$m$ fading channels. Apart from
above, we have shown the impact of power allocation factor ($\alpha$)
and parameter $m$ on the region of secondary spectrum access, conventionally
defined as critical radius for the secondary system. A comparison
between theoretical and simulated results is also presented to corroborate
the theoretical results obtained in this paper. \end{abstract}
\begin{IEEEkeywords}
Nakagami-$m$ fading, cognitive radio, spectrum sharing, decode and
forward relaying, cooperative communication. 
\end{IEEEkeywords}

\section{Introduction}

\IEEEPARstart{ C}{ooperative} spectrum sharing (CSS) have attracted
a great deal of attention among researchers in the past few years
due to its dual utilization of cooperative diversity for reliable
communication and cognitive abilities to utilize the spectrum more
efficiently \cite{key-1,key-2}. The concept of CSS, to employ the
secondary transmitter (ST) as a relay to forward the information of
the primary system and get spectrum access in exchange, can be utilized
in cellular and ad-hoc networks \cite{key-1}-\cite{key-3}.

Considerable work has been done to validate the performance of CSS
protocols in Rayleigh faded channels, however to the best of our knowledge
very few literature is publicly available to demonstrate the performance
of these protocols on Nakagami faded channels. Many experimental works
show that Nakagami distribution, as compared to Rayleigh distribution,
is often more accurate for modeling the urban multipath channels \cite{key-4}.
Although Rayleigh fading models are frequently utilized in modeling
the non light-of-sight channels however it is better fit for the signals
propagating within small areas, as it does not gauge for large-scale
propagation effects like shadowing by buildings, bridges and other
obstructions which are typically encountered in mobile communication
channel. Hence, Nakagami fading models are usually preferred in modeling
long distance fading effects, specifically with respect to mobile
communications \cite{key-4}. 

Moreover, by tuning the fading severity parameter, m , Nakagami-m
fading can be used to represent a wider class of fading channel conditions.
For instance, m=1 represents Rayleigh fading whereas m = 0.5 represents
one-sided Gaussian fading \cite{key-5}. 

The authors in \cite{key-6} have analyzed the performance of classical
decode and forward (DF) cooperative communications over Nakagami-$m$
fading channels. They have measured the performance in terms of symbol
error rate (SER) for different modulation schemes. By varying the
parameters of the fading channel the authors are able to enhance the
cooperation performance between primary and secondary system. In \cite{key-7,key-8}
the outage performance of an underlay system with cognitive decode
and forward (DF) and amplify and forward (AF) relaying schemes has
been investigated. The authors in their system model have used relays
for transmitting secondary system's data. Secondary transmitter and
its relay limits their transmit power so that the interference on
the primary system do not exceed a certain threshold.

Compared to the previous work proposed in the literature on Nakagami
fading channels, in this paper we have considered an overlay model
in which there is no limitation on the secondary transmit power. On
a contrary, depending on the power allocation factor, $\alpha$, the
performance of the primary and secondary system may increase with
an increase in the secondary transmit power. This paper can also be
seen as an extension of the work done for Rayleigh fading channels
in \cite{key-1,key-2}, however, in the proposed work the results
have been obtained for independent Nakagami-$m$ fading channels.
Furthermore, unlike \cite{key-1}, we believe that cognitive system
is going to be an advance system that utilizes the multiple antenna
functionality \cite{key-3} such as IEEE 802.11n, IEEE 802.16m or
3GPP LTE - Advanced \cite{key-9}. Hence, in the proposed work, it
is assumed that ST is equipped with multiple antennas.

In our system, when the target rate of the primary system drops below
a particular threshold ($\mathcal{R}_{pt}$), it seeks cooperation
from the neighboring terminals. The secondary system which disguises
itself as relay, cooperates with the primary system, promising better
performance to primary system in exchange for the spectrum access
in the operating frequency band of primary system. Once ST is confirmed
as a relay, spectrum access for secondary system is obtained by adopting
the following two-phase transmission protocol. In phase 1, the data
broadcasted from primary transmitter (PT) is received by primary receiver
(PR), secondary transmitter (ST) and secondary receiver (SR). The
data received at PR in phases 1 and 2 is decoded using maximum ratio
combining (MRC) to get the desired data, considering secondary data
as noise. At SR, after successful decoding of primary signal in phase
1, the interference component can be canceled out in phase 2 to obtain
the desired secondary data \cite{key-1,key-2}%
\footnote{Interested readers may refer to \cite{key-1} for further details
on the control protocol. %
}. Our proposed model has been quantified by obtaining closed form
expressions for the outage probability. Apart from above, we have
calculated the critical region of ST which helps in determining the
maximum distance within which ST can achieve spectrum access.

\section{System Model}

The system model for the $1^{\mathrm{st}}$ and $2^{\mathrm{nd}}$
phase is shown in fig 1. 
\begin{figure}[tbh]
\centering\includegraphics[width=3.5in]{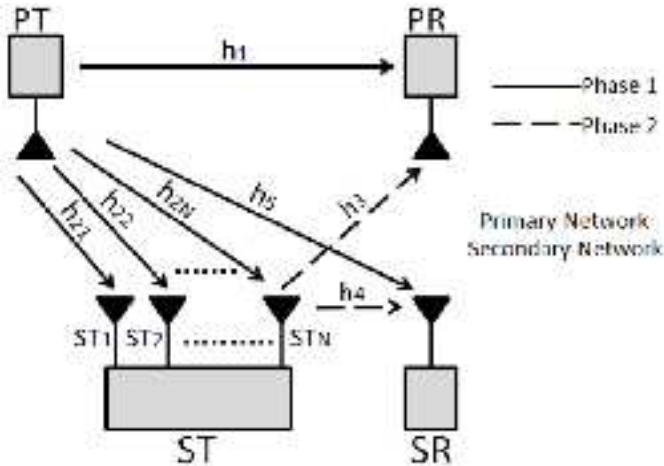}\caption{System Model}
\end{figure}

The system consists of PT, PR, ST and SR. The channel between all
the links, i.e. PT-PR, PT-$\mathrm{ST_{1}}$, PT-$\mathrm{ST_{2}}$...PT-$\mathrm{ST_{N}}$,
$\mathrm{ST_{n}}$-PR, $\mathrm{ST_{n}}$-SR, where n is $n^{th}$
antenna selected randomly, and PT-SR are described by Nakagami-$m$
distribution with channel coefficients $h_{1},\: h_{(21)},\: h_{(22)},.......,\: h_{(2N)},\: h_{3},\: h_{4}$
and $h_{5}$ respectively. The probability density function (PDF)
of a Nakagami random variable $\gamma=|h|^{2}$ is given by 

\[
p_{\gamma}(h)=\frac{2m^{m}h^{2m-1}}{\Gamma(m)\Omega^{m}}e^{-\frac{mh^{2}}{\Omega}},
\]
where $\Omega=E\{\gamma^{2}\}$ is the variance of $\gamma$, $m$
is the Nakagami fading figure and $\Gamma(.)$ is the Gamma function.
Generally, when $m=1$ the above PDF reduces to the PDF of well-known
Rayleigh fading model. For $0.5\leq m<1$, the fading is Nakagami
which is more severe than that of Rayleigh fading.

The parameter $\Omega_{i}=d_{i}^{-k}$ where $k$ is the path loss
component and $d_{i}$ is the normalized distance between the respective
transmitters and receivers. This normalization is done with respect
to distance between PT and PR, i.e. $d_{1}=1$. Primary and secondary
signals are denoted by $x_{p}$ and $x_{s}$ respectively, with zero
mean and $E\{|x_{p}|^{2}\}=1,\: E\{|x_{s}|^{2}\}=1$. $\mathcal{R}_{pt}$
and $\mathcal{R}_{st}$ are the target rates for $x_{p}$ and $x_{s}$
respectively. We denote the transmit power at PT and ST as $P_{p}$
and $P_{s}$, respectively. The additive white Gaussian noise (AWGN)
at each receiver is denoted by $n_{ij}\sim\mathit{\mathcal{CN}}(0,\sigma^{2})$
where $i\epsilon\{1,2\}$ represents the transmission phase and $j\epsilon\{1,(21),(22),...,(2N),..,5\}$
represents the respective channel link, assumed to have identical
variance $\sigma^{^{2}}$. In the following sections we will analyze
the performance of cooperative spectrum sharing based on DF protocol
under Nakagami-$m$ fading channels.

\section{Outage Performance of Primary System}

In phase 1, PT will broadcast the signal $x_{p}$. This signal is
overheard by PR, $\mathrm{ST_{1}}$, $\mathrm{ST_{2}}$, ..., $\mathrm{ST_{N}}$
and SR. The received signal at PR is denoted by $y_{pr_{1}}$, which
is given by 

\[
y_{pr_{1}}=\sqrt{P_{p}}h_{1}x_{p}+n_{11},
\]
where, $n_{11}\sim\mathit{\mathcal{CN}}(0,\sigma^{2})$. The received
signal at ST is denoted by 

\[
\mathbf{y}_{\mathbf{\mathrm{ST}}}=\mathbf{h}_{df}x_{p}+n,
\]
where, $\mathbf{y}_{\mathbf{\mathrm{ST}}}=[y_{st}^{\{21\}}\; y_{st}^{\{22\}}...\; y_{st}^{\{2N\}}]^{T}$.
$y_{st}^{\{21\}},\; y_{st}^{\{22\}},...\; y_{st}^{\{2N\}}$ denote
the signals coming from channel $h_{(21)},\; h_{(22)},...,\; h_{(2N)}$,
respectively. Also, 
\[
\mathbf{h}_{df}=[\sqrt{P_{p}}h_{(21)}\;\sqrt{P_{p}}h_{(22)}....\;\sqrt{P_{p}}h_{(2N)}]^{T}
\]
and $n=[n_{1(21)\;}n_{1(22)}\;....n_{1(2N)}]^{T}.$ The signals thus
received at ST is decoded for $x_{p}$. The rate at ST is given by,
\begin{gather*}
\mathcal{R}_{ST}=\frac{1}{2}\mathrm{log_{2}}\left[1+\frac{P_{p}\left(\left|h_{(21)}\right|^{2}+\left|h_{(22)}\right|^{2}+...+\left|h_{(2N)}\right|^{2}\right)}{\sigma^{2}}\right]
\end{gather*}
\begin{equation}
\mathcal{R}_{ST}=\frac{1}{2}\mathrm{log_{2}}\left[1+\frac{P_{p}\left(\left\Vert h\right\Vert ^{2}\right)}{\sigma^{2}}\right].
\end{equation}
where $||h||^{2}=|h_{(21)}|^{2}+|h_{(22)}|^{2}+...+|h_{(2N)}|^{2},$
the factor $\frac{1}{2}$ in the above equation accounts for the fact
that the transmission is being divided into two phases. In phase 2,
if $x_{p}$ is decoded successfully, ST will transmit $x_{p}$ along
with its own data $x_{s}$. The signals received at PR is given by 

\[
\mathit{y}_{pr_{2}}=\mathbf{g_{\mathit{df}}}\mathbf{v}_{st}+n_{23},
\]
where 
\[
\mathbf{g}_{df}=\begin{bmatrix}\sqrt{\alpha P_{s}}h_{3} & \sqrt{(1-\alpha)P_{s}}h_{3}\end{bmatrix},
\]

\[
\mathbf{v}_{st}=\begin{bmatrix}x_{p} & x_{s}\end{bmatrix}^{T}
\]
 and $n_{23}\sim\mathit{\mathcal{CN}}(0,\sigma^{2}).$ The signals
in phase 1 and 2, $y_{pr_{1}}$ and $y_{pr_{2}}$, are then combined
at PR using MRC. The achieved rate is then derived as in \cite{key-1}
is given by 
\begin{equation}
\mathcal{R}_{p}=\frac{1}{2}\mathrm{log_{2}}\left(1+\frac{P_{p}\gamma_{1}}{\sigma^{2}}+\frac{\alpha P_{s}\gamma_{3}}{\left(1-\alpha\right)P_{s}\gamma_{3}+\sigma^{2}}\right).
\end{equation}
On the other hand if ST is not able to decode $x_{p}$ in phase 1
then it will not transmit in phase 2. In such a case PR can still
receive $x_{p}$ through a direct link from PT to PR with achievable
rate of 

\[
\mathcal{R}_{pd}=\mathrm{log_{2}}\left(1+\frac{P_{p}\gamma_{1}}{\sigma^{2}}\right).
\]
Thus, the outage probability of the primary signal transmission with
target rate $\mathcal{R}_{pt}$ is given as 
\begin{alignat}{1}
F_{op} & =1-P_{r}\{\mathcal{R}_{ST}>\mathcal{R}_{pt}\}P_{r}\{\mathcal{R}_{p}>\mathcal{R}_{pt}\}\nonumber \\
 & -P_{r}\{\mathcal{R}_{ST}<\mathcal{R}_{pt}\}P_{r}\{\frac{1}{2}\mathcal{R}_{pd}>\mathcal{R}_{pt}\}.
\end{alignat}
Assuming $P_{s}>>\sigma^{2}$ as in \cite{key-1}, \cite{key-2} we
obtain 
\begin{align}
 & P_{r}\{\mathcal{R}_{p}>\mathcal{R}_{pt}\}\nonumber \\
 & =\begin{cases}
\frac{1}{\Gamma(m)}\left[\Gamma(m)-\Gamma\left(m,\frac{m\sigma^{2}}{\Omega_{1}p_{p}}\left(\rho_{1}-\frac{\alpha}{1-\alpha}\right)\right)\right] & 0\leq\alpha<\hat{\alpha}\\
1 & \hat{\alpha}\leq\alpha<1
\end{cases},
\end{align}
where $\rho_{1}=2^{2\mathcal{R}_{pt}}-1$, $\hat{\alpha}=\frac{\rho_{1}}{\rho_{1}+1}$,
$\Gamma(.)$ is the Gamma function and $\Gamma(.,.)$ indicate the
incomplete Gamma function. 
\begin{alignat}{1}
P_{r}\{\mathcal{R}_{ST}<\mathcal{R}_{pt}\} & =\frac{1}{\Gamma(Nm)}\left[\Gamma\left(Nm,\frac{m\sigma^{2}}{\Omega_{2}p_{p}}\rho_{1}\right)\right].\\
P_{r}\{\mathcal{R}_{ST}>\mathcal{R}_{pt}\}= & \frac{1}{\Gamma(Nm)}\nonumber \\
 & \left[\Gamma(Nm)-\Gamma\left(Nm,\frac{m\sigma^{2}}{\Omega_{2}p_{p}}\rho_{1}\right)\right].\\
Pr\{\frac{1}{2}\mathcal{R}_{pd}>\mathcal{R}_{pt}\} & =\frac{1}{\Gamma(m)}\left[\Gamma(m)-\Gamma\left(m,\frac{m\sigma^{2}}{\Omega_{1}p_{p}}\rho_{1}\right)\right].
\end{alignat}
Substituting (4), (5), (6) and (7) in (3). We get, 
\begin{alignat}{1}
F_{op} & =\begin{cases}
F_{op_{1}}, & 0\leq\alpha<\hat{\alpha}\\
F_{op_{2}}, & \hat{\alpha}\leq\alpha<1
\end{cases}
\end{alignat}
\[
F_{op_{1}}=1-\frac{1}{\Gamma(Nm)}\left[\Gamma(Nm)-\Gamma\left(Nm,\frac{m\sigma^{2}}{\Omega_{2}p_{p}}\rho_{1}\right)\right]
\]
\[
\frac{1}{\Gamma(m)}\left[\Gamma(m)-\Gamma\left(m,\frac{m\sigma^{2}}{\Omega_{1}p_{p}}\left(\rho_{1}-\frac{\alpha}{1-\alpha}\right)\right)\right]-\frac{1}{\Gamma(Nm)}
\]
\begin{equation}
\left[\Gamma\left(Nm,\frac{m\sigma^{2}}{\Omega_{2}p_{p}}\rho_{1}\right)\right]\frac{1}{\Gamma(m)}\left[\Gamma(m)-\Gamma\left(m,\frac{m\sigma^{2}}{\Omega_{1}p_{p}}\rho_{1}\right)\right]
\end{equation}

\[
F_{op_{2}}=1-\frac{1}{\Gamma(Nm)}\left[\Gamma(Nm)-\Gamma\left(Nm,\frac{m\sigma^{2}}{\Omega_{2}p_{p}}\rho_{1}\right)\right]
\]
\[
\frac{1}{\Gamma(Nm)}\left[\Gamma\left(Nm,\frac{m\sigma^{2}}{\Omega_{2}p_{p}}\rho_{1}\right)\right]
\]
\begin{equation}
\frac{1}{\Gamma(m)}\left[\Gamma(m)-\Gamma\left(m,\frac{m\sigma^{2}}{\Omega_{1}p_{p}}\rho_{1}\right)\right]
\end{equation}

\section{Region for secondary spectrum access}

In this section, we are going to define the region, within which the
secondary system can access primary's spectrum without compromising
the performance of primary system. This region has been conventionally
defined as critical radius in \cite{key-1}. To calculate a critical
region for such a system, the outage probability of primary system
with cooperation i.e. $F_{op}$, must be less than the outage probability
without cooperation, i.e. $F_{op}\leq P_{d}$. The outage probability
of direct transmission (without cooperation) is given as 
\begin{alignat}{1}
P_{d} & =P_{r}\{\mathcal{R}_{pd}<\mathcal{R}_{pt}\}=\frac{1}{\Gamma(m)}\left[\Gamma\left(m,\frac{m\sigma^{2}}{\Omega_{1}p_{p}}\rho_{2}\right)\right],
\end{alignat}
where $\rho_{2}=2^{\mathcal{R}_{pt}}-1$ and $\Gamma(.,.)$ indicate
the incomplete Gamma function. From (9), (10), we can observe that
$F_{op}$ not only depends on the $\Omega$ but it also varies with
change in the value of $\alpha$. Therefore, there are two cases which
describes the successful spectrum access of the secondary system.
i.e for $\Omega_{2}\leq\widetilde{\Omega}_{2}$ and $\alpha>\widetilde{\alpha}$.
The theoretical values of $\widetilde{\Omega}_{2}$ after solving
$F_{op_{2}}\leq P_{d}$ is given as below 
\begin{equation}
\Omega_{2}\leq\widetilde{\Omega_{2}}=\left[\frac{P_{p}}{m\rho_{1}\sigma^{2}}\Gamma^{-1}\left(Nm,\frac{\Gamma(m,\frac{m\sigma^{2}}{P_{p}}\rho_{2})}{\Gamma(m,\frac{m\sigma^{2}}{P_{p}}\rho_{1})}\right)\right]^{-1},
\end{equation}
where $\Gamma^{-1}(\text{.},\text{.})$ indicate the inverse incomplete
Gamma function. The $\widetilde{\alpha}$ for $P_{d}\geq F_{op_{1}}$
is given as 
\begin{equation}
\alpha\geq\widetilde{\alpha}=\frac{\rho_{1}-\chi}{1+\rho_{1}-\chi},
\end{equation}
where 
\[
\chi=\left[\frac{\Omega_{1}P_{p}}{m\sigma^{2}}\Gamma^{-1}\left(m,\varphi\right)\right]
\]
 and 
\[
\varphi=\frac{\Gamma\left(m,\frac{m\sigma^{2}}{P_{p}}\rho_{2}\right)-\Gamma\left(m,\frac{m\sigma^{2}}{P_{p}}\rho_{1}\right)\Gamma\left(Nm,\frac{m\sigma^{2}}{\Omega_{2}P_{p}}\rho_{1}\right)}{1-\Gamma\left(Nm,\frac{m\sigma^{2}}{P_{p}\Omega_{2}}\rho_{1}\right)}.
\]
 We can note that for $m=1$ (12), (13) reduces to the results given
in \cite{key-1,key-2} for Rayleigh flat fading.

\section{Outage Performance of Secondary System}

In phase 1, received signals at secondary receiver is given by 
\[
y_{sr_{1}}=\sqrt{P_{p}}h_{5}x_{p}+n_{15},
\]
where $n_{15}\sim\mathit{\mathcal{CN}}(0,\sigma^{2})$. The rate at
SR for the direct transmission from PT is given by 
\begin{equation}
\mathcal{R}_{sd}=\frac{1}{2}\mathrm{log_{2}}\left[1+\frac{P_{p}\gamma_{5}}{\sigma^{2}}\right].
\end{equation}
At SR, an estimate of $x_{p}$ is obtained as 

\[
\widehat{x_{p}}=\frac{y_{sr_{1}}}{\sqrt{P_{p}}h_{5}}=x_{p}+\frac{n_{15}}{\sqrt{P_{p}}h_{5}}.
\]
The achievable rate at ST is given in (1). In phase 2, signal received
at SR is given by 

\[
y_{sr_{2}}=\mathbf{h}_{s}\mathbf{v}_{st}+n_{24},
\]
where 
\[
\mathbf{h}_{s}=\begin{bmatrix}\sqrt{\alpha P_{s}}h_{4} & \sqrt{(1-\alpha)P_{s}}h_{4}\end{bmatrix},
\]
and $n_{24}\sim\mathit{\mathcal{CN}}(0,\sigma^{2})$ is the AWGN.
The estimate $\widehat{x_{p}}$ is used to cancel the interference
component, to obtain 
\[
\widehat{y}_{sr_{2}}=\sqrt{(1-\alpha)P_{s}}h_{4}x_{s}+n_{24}.
\]
The achieved rate between ST and SR, conditioned on successful decoding
of $x_{p}$ at both ST and SR in the first phase, is given by 
\begin{equation}
\mathcal{R}_{s}=\frac{1}{2}\mathrm{log_{2}}\left[1+\frac{(1-\alpha)P_{s}\gamma_{4}}{\sigma^{2}}\right].
\end{equation}
Outage is declared if ST and SR are not able to decode $x_{p}$, and
therefore the outage probability of the secondary signal transmission
with target rate $\mathcal{R}_{st}$ is given as

\begin{eqnarray}
F_{os} & = & 1-P_{r}\{\mathcal{R}_{ST}>\mathcal{R}_{pt}\}\nonumber \\
 &  & P_{r}\{\mathcal{R}_{sd}>\mathcal{R}_{pt}\}P_{r}\{\mathcal{R}_{s}>\mathcal{R}_{st}\}.
\end{eqnarray}

\begin{align}
P_{r}\{\mathcal{R}_{s} & >\mathcal{R}_{st}\}=\nonumber \\
 & \frac{1}{\Gamma(m)}\left[\Gamma(m)-\Gamma\left(m,\frac{m\sigma^{2}}{\Omega_{4}\left(1-\alpha\right)P_{s}}\rho_{3}\right)\right],
\end{align}

\begin{alignat}{1}
 & Pr\{\mathcal{R}_{sd}>\mathcal{R}_{pt}\}=\frac{1}{\Gamma(m)}\left[\Gamma(m)-\Gamma\left(m,\frac{m\sigma^{2}}{\Omega_{5}P_{p}}\rho_{1}\right)\right],
\end{alignat}
where $\rho_{3}=2^{2R_{st}}-1$. Substituting (6), (17) and (18) in
(16) we get the outage probability as 
\begin{alignat}{1}
F_{os}= & 1-\frac{1}{\Gamma(Nm)}\left[\Gamma(Nm)-\Gamma\left(Nm,\frac{m\sigma^{2}}{\Omega_{2}P_{p}}\rho_{1}\right)\right]\nonumber \\
 & \frac{1}{\Gamma(m)}\left[\Gamma(m)-\Gamma\left(m,\frac{m\sigma^{2}}{\Omega_{5}P_{p}}\rho_{1}\right)\right]\nonumber \\
 & \frac{1}{\Gamma(m)}\left[\Gamma(m)-\Gamma\left(m,\frac{m\sigma^{2}}{\Omega_{4}\left(1-\alpha\right)P_{s}}\rho_{3}\right)\right].
\end{alignat}

\section{Simulation Results and Discussions}

In this section, we discuss the performance of a cooperative spectrum
sharing protocol for Nakagami-m fading. Target rates of primary as
well as secondary systems are chosen to be $\mathcal{R}_{pt}=\mathcal{R}_{st}=1$.
The value of $m$ is taken as $m=0.7$, which measures the depth of
fading envelope. PT, ST, SR, PR nodes are assumed to be collinear
as in \cite{key-1}, \cite{key-10}. The node ST is equipped with
N antennas. For simulation, we have taken the value of N=2 and N=4.
The distance between PT and PR is normalized and taken as $d_{1}=1$.
The distance between PT and ST is denoted by $d_{2}$ and the respective
distances between different nodes is calculated in terms of $d_{2}$
. The distance between ST and PR is $d_{3}=|1-d_{2}|$, ST to SR and
PT to SR is $d_{4}=d_{5}=d_{2}/2$. We have taken $\frac{P_{p}}{\sigma^{2}}=20\mathrm{dB}$
and $\frac{P_{s}}{\sigma^{2}}=30\mathrm{dB}$. The $k=4$, is the
path loss component.

Fig. 2 shows the outage probability performance of primary system
w.r.t. the power allocating factor $\alpha$ for different values
of $d_{2}=\{0.8,\;1.5,\;3.25\}$ for N=4 and $d_{2}=\{0.8,\;1.5,\;2.625\}$
for N=2. These set values of $d_{2}$ are calculated from $\Omega_{2}=d_{2}^{-k}$,
where the last values in both the sets are the critical values calculated
from (12). It can be seen from the figure that as we increase the
value of $\alpha$ the outage probability tends to decrease. 
\begin{figure}[tb]
\centering\includegraphics[width=3.5in]{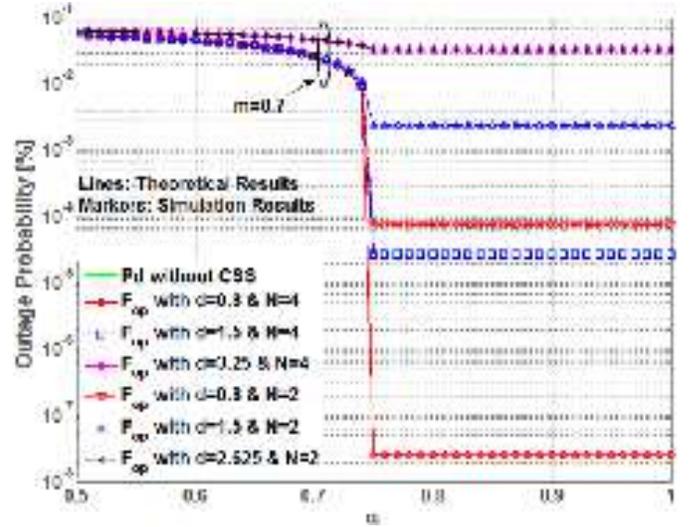}\caption{Outage Probability of Primary System}
\end{figure}

For a particular value of $d_{2}$ between PT and ST, when $\alpha>\tilde{\alpha}$
the outage probability drops below the outage probability of direct
link and spectrum access can be achieved by secondary system. As we
increase $\alpha\geq\hat{\alpha}$ then for a particular $d_{2}$
between PT and ST much lower outage probability can be achieved by
the primary system. The outage probability performance of the system
under Nakagami fading reduces to Rayleigh fading for $m=1$.%
\footnote{In fig. 2, theoretical results are plotted by assuming $P_{s}>>\sigma^{2}$,
however, for small values of $P_{s}$ the approximation $P_{s}>>\sigma^{2}$
might not hold and there would be a slight gap between the simulation
and theoretical results.%
}

Fig. 3 shows reasonably good outage probability of secondary system
w.r.t. $\alpha$. The theoretical results are exactly matching with
the simulation results, authenticating the analytical results obtained
for the outage probability of secondary system. We can observe from
figure that the outage probability has a constant value for almost
all values of $\alpha$ and tends to 1 as $\alpha\rightarrow1$. 
\begin{figure}[tbh]
\centering\includegraphics[width=3.5in]{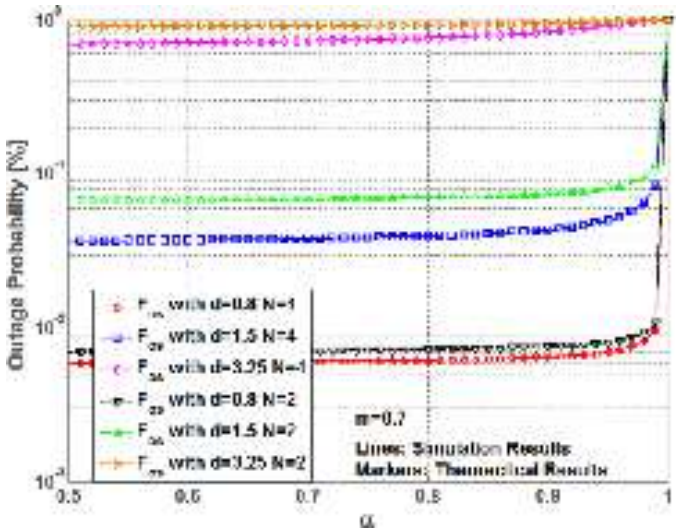}
\caption{Outage Probability of Secondary System}
\end{figure}
\begin{figure}[tb]
\centering\includegraphics[width=3.5in]{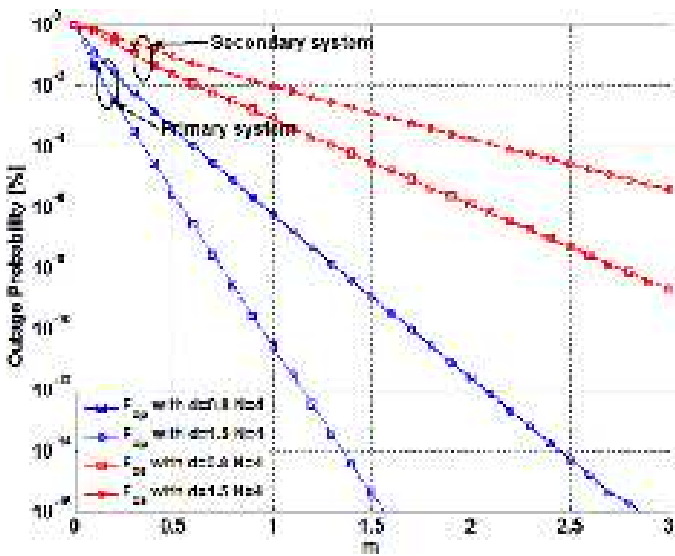}
\caption{Outage Probability of primary and secondary system w.r.t fading parameter
m}
\end{figure}

Fig. 4 shows the outage probability of the primary as well as the
secondary system w.r.t the fading coefficient $m$. It can be observed
from the figure that as the value of $m$ increases i.e. the fading
effect of the channel decreases, the outage probability of the overall
system decreases, which is quite obvious from the fact that as there
is no fading in the channel the data can be transmitted smoothly and
efficiently to the destination. From fig. 3 and fig. 4 it can be inferred
that there is good agreement between theoretical and simulating results
thus validating the analysis done in this paper.

\section{Conclusions}

In this paper, we analyzed the performance of cooperative spectrum
sharing scheme over Nakagami-$m$ fading. A cognitive relay, equipped
with multiple antennas decodes the message from primary transmitter
and forwards, by means of DF relaying, it to the destination by randomly
selecting one antenna in order to achieve the target rate of primary
system, getting the spectrum access for secondary system in exchange.
It was shown that, even in presence of Nakagami-$m$ fading CSS protocol
with multiple antennas at ST can help in considerable improvement
in the performance of primary system. From above observations, we
can conclude that as the value of $m$ increases the severity of fading
decreases and performance of outage probability improves. The excellent
agreement between the simulated results and the analytically obtained
closed form expressions authenticates the theoretical analysis presented
in this paper.

\end{document}